\documentclass[twocolumn,showpacs,preprintnumbers,amsmath,amssymb]{revtex4}

\usepackage{graphicx}
\usepackage{dcolumn}
\usepackage{bm}

\newcommand{\RR}{\mathbb{R}}
\newcommand{\hhf}{{\scriptstyle{{\frac{1}{2}}}}}

\def\mr{{\bf r}}
\def\mb{{\bf b}}

\begin{document}

\preprint{APS/123-QED}

\title{Dynamical Systems to Monitor Complex Networks in Continuous Time}

\author{Peter Grindrod}
 \altaffiliation[from 1st Sept:~]  
{Mathematical Institute, 
University of Oxford, OX1 3LB, UK}
\affiliation{Department of Mathematics, University of Reading, Reading 
RG6 6AX, UK}
\author{Desmond J. Higham}%
 \email{d.j.higham@strath.ac.uk}
\affiliation{%
Department of Mathematics and Statistics
University of Strathclyde, 
Glasgow, G1 1XH, UK
}%

\date{\today}

\begin{abstract}
In many settings it is appropriate to treat the evolution of pairwise interactions over continuous time.
We show that new Katz-style centrality measures can be derived in this context 
via solutions to a nonautonomous ODE driven by the network dynamics.
This allows us to identify and track, at any resolution, the most influential nodes in terms of
broadcasting and receiving information through time dependent links.
In addition to the classical notion of attenuation across edges used in 
the static Katz centrality measure, the 
ODE also allows for attenuation over time, so that real time ``running measures'' can be 
computed.     
With regard to computational efficiency, 
we explain why 
it is cheaper to track 
good receivers of information than good broadcasters.
We illustrate the new measures on a large scale voice call network, where 
key features are discovered that are not evident from snapshots or aggregates.
\end{abstract}

\maketitle


Many systems of interacting agents involve 
transient edges, and this time-dependency 
raises a variety of challenges \cite{HS2013}.
Currently, the dominant paradigm is to 
consider activity over time-windows, and hence to  
deal with network snapshots over discrete time. 
This approach is useful for modeling both the evolution of   
the underlying connectivity and the dynamics of processes over the network
\cite{Ba13,La11,St11}.
The idea of 
viewing the network as a 
time-ordered sequence of sparse adjacency matrices 
is also extremely convenient for data-driven analysis
\cite{GHPE11,GH2013,TMMLN10,LSS13,Mucha2010,tang-2009}.
More generally, abstracting from time to any 
countable index representing possible levels in a hierarchy 
(such as gene/mRNA/protein)
we arrive at 
multiplex networks 
\cite{GDD2013},
which can be usefully viewed as multidimensional tensors
\cite{DKA11}.

However, 
in many settings 
using 
a time-ordered sequence of network snapshots 
forces us to make awkward compromises. 
On one hand, by taking the time window too large we lose the ability to reproduce
high-frequency transient behaviour, where an edge switches on and off multiple times in the space of a single window.
On the other hand, time points that are too finely spaced 
can lead to redundant, empty windows that waste computational effort. 
Very short time windows may also lead to a false impression of
accuracy---for remote interactions such as emails and tweets, although there is an unambiguous and instantaneous time
at which information was sent, the time at which the receiver (a) digests and then, possibly,
(b) acts upon the information is generally not recordable.

In this work we will show that it is 
possible to extract insight directly from an 
adjacency matrix defined as a 
function of continuous time.
We focus on the computationally-motivated task of 
monitoring the influence of nodes in a given network. 
However, we note that 
the ODE setting developed here also has potential for high-level analysis and prediction,
since the tools of dynamical systems
become available when 
network models are combined with the centrality evolution; for example, 
via a law of social balance 
 \cite{Marvel11,TVD2013} or triadic closure 
 \cite{GHPIM12}.
Hence, there is potential to study the evolution of centrality under various 
network ``laws of motion'' and to compare predictions from competing models.

For a fixed set of $N$ nodes, we let 
$A(t)$ denote the state of the network at time $t$.
In the case of voice call data, it is natural for the entries in the adjacency matrix 
to be discontinuous, binary-valued, functions of time, switching on to 1 and off to 0 when an exchange starts and finishes,
respectively.
So in this case each 
$A(t) \in \RR^{N \times N}$ is binary and symmetric 
with $A_{ij}(t)  = A_{ji} (t) = 1$ if and only if nodes
$i$ and $j$ are in communication at time $t$.
(To be concrete, and without loss of generality, we will assume 
continuity from the right.)
For social interactions such as emails and tweets, this methodology allows us to
spike the relevant edge when a message is sent out,
and then, as $t$ increases, let the value to decay towards zero---this reflects the
fact that these exchanges have a natural delay, and also that messages lose their impact and visibility over time.
In this case $A(t)$ is generally unsymmetric, taking values between zero and one.

Our aim is now to generate a real-valued, unsymmetric 
\emph{dynamic communicability matrix} 
$S(t) \in \RR^{N \times N}$ such that 
$S(t)_{ij}$  quantifies the ability of  
node $i$ to communicate with node $j$ up to time $t$, that is, by making use of edges that 
have appeared at time $t$ or earlier.
We propose that $S(t)$ is  updated over a small time interval $\delta t$ according to
\begin{equation}
\label{eq:Std}
S(t+\delta t ) = \left(I+e^{-b \delta t} S(t)\right) \left(I-a A(t+\delta t)\right)^{-\delta t}-I,
\end{equation}
with $S(0) = 0$.
Here, $I \in \RR^{N \times N}$ denotes the identity matrix, 
and $a, b > 0$ are parameters. To justify this new iteration 
and explain the role of the parameters, suppose first that 
$\delta t = 1$, and we are therefore given access to the adjacency matrix only on a fixed 
grid of timepoints.
In the extreme case of a single timepoint, (\ref{eq:Std}) reduces to 
$S(1) + I = (I-a A(1))^{-1}$.
This corresponds to the classical static network analysis 
measure of Katz 
\cite{Katz53}; for $i \neq j$, 
$S(1)_{ij}$ counts the total number of distinct walks through the network from  node $i$ to node $j$, 
with walks of length $k$ downweighted by $a^k$. The attenuation factor $a$
may be interpreted as the probability that a message successfully traverses an edge.
With multiple time points and $b = 0$ in 
(\ref{eq:Std})
we have 
\begin{eqnarray}
S(N) + I &=& 
 \left( 
I-a A(1)
\right)^{-1}
 \left( 
I-a A(2)
\right)^{-1}
\cdots
       \nonumber 
 \\
&& 
\mbox{}
 ~~~~~~~~
 ~~~~~~~~
 ~~~~~~~~
 \left( 
I-a A(N)
\right)^{-1}.
 \label{eq:disc}
\end{eqnarray}
It was shown in 
\cite{GHPE11}
that this iteration generalizes the Katz centrality 
measure to the case of a time-ordered network sequence. Here
$S(N)_{ij}$ records an edge-downweighted count of the total number of distinct \emph{dynamic  walks} through the network,
where a dynamic walk is any traversal that respects the time-ordering of the edges.
We note that the same idea was later used to define concept of accessibility \cite{LSS13}.
The general case of (\ref{eq:Std}) with $ b > 0$ and $\delta  t = 1$ was introduced in 
\cite{GH2013}, where it was shown that 
the parameter $b$ plays the role of downweighting for time---walks that began $k$ time units ago are scaled by 
$e^{-bk}$.  This temporal scaling is designed to take account of the fact that 
older messages generally have less current relevance.

A key step in writing down the new iteration 
(\ref{eq:Std}) was to devise a scaling under which the 
$\delta t  \to 0$ limit is meaningful.
Our choice of a $\delta t$-dependent power in the matrix products can be explained by 
focusing on the $b = 0$ case over a short time interval.
If we refine from the single interval 
 $[t,t+\delta t]$
 to the pair of intervals 
 $[t,t+\delta t/2]$
 and 
 $[t+\delta t/2, t + \delta t]$ and consider the scenario where $A(t)$ does not change, the 
simple identity 
\[
\left(I-a A(t)\right)^{-\hhf \delta t} 
\left(I-a A(t)\right)^{- \hhf \delta t} 
=
\left(I-a A(t)\right)^{-\delta t} 
\]
shows that 
our scaling produces consistent results.

For convenience, we will let $U(t) = I + S(t)$ in taking the 
$\delta t  \to 0$ limit.
We then write 
(\ref{eq:Std}) in the form 
\begin{eqnarray*}
U(t+\delta t ) &=& \left(I+e^{-b \delta t} \left(  U(t) - I \right) )\right) 
 \\
 && \mbox{}
 \times 
      \exp \left(
                - \delta t \log\left(I-a A(t+\delta t)\right) \right),
\end{eqnarray*}
where $\exp$ and $\log$ denote the matrix exponential and logarithm respectively, and we have used the 
identities $H = e^{\log H}$ 
and
$\log (H^\alpha) = \alpha \log H$ for $-1 \le \alpha \le 1$; 
see, for example, \cite{HigFM}.
Expanding and 
taking the 
limit $\delta t  \to 0$ we then arrive at the matrix ODE
\begin{equation}
\label{eq:Uode}
U'(t) = - b \left( U(t) - I \right) 
            - U(t)  \log \left( I - a A(t) \right), 
\end{equation}
with $U(0) = I$.

In order to focus on nodal properties, we can take row and column sums in the matrix $U(t)$ to define the
\emph{dynamic broadcast} and 
\emph{dynamic receive} vectors
\begin{equation}
\mb(t) = U(t) \mathbf{1} 
 \quad
 \mathrm{and} 
\quad 
\mr(t) = U(t)^T \mathbf{1},
\label{eq:brvec}
\end{equation}
where $\mathbf{1} \in \RR^{N}$ denotes the vector of ones. Here, 
$\mb_i(t)$ and 
$\mr_i(t)$ measure the current propensity for node $i$ to
have  
broadcasted and received information, respectively, across the dynamic network, under our 
assumptions that longer and older walks have less relevance.

It is interesting to note 
 from
(\ref{eq:Uode}) that 
the receive centrality satisfies its own vector-valued ODE
\[
\mr'(t) = - b \left( r(t) - \mathbf{1} \right) 
            -  \left( \log \left( I - a A(t) \right) \right)^T \mr(t), 
\]
with $\mr(0) =  \mathbf{1}$, which is a factor of $N$ smaller in dimension than 
(\ref{eq:Uode})
and hence cheaper to simulate.
By contrast, it is not possible to 
disentangle the 
broadcast centrality in this way. Intuitively, this difference arises because the 
node-based receive vector 
$\mr(t)$ keeps track of the overall level of information flowing into each node, and this can be 
propagated forward in time as new links become available. 
	However, the broadcast vector 
	$\mb(t)$ keeps track of information that has flowed out of each node---we do not record where the information currently resides and hence we cannot 
	update based on $\mb(t)$ alone.
	So, with this methodology, 
	real time updating of the receive centrality is 
	fundamentally simpler than 
	real time updating of the broadcast centrality.

	The 
	choice of 
	downweighting parameters, $a$ and $b$, clearly plays an important role, and should depend to a large extent on the 
	nature of the interactions represented by $A(t)$.
	The temporal downweighting parameter, $b$, has the natural interpretation that walks starting $t$ time units ago are 
	downweighted by 
	$e^{- b t}$, so $b$ allows us summarize the rate at which news goes stale.
	Hence $b$ could be calibrated by studying the typical half-life of a link 
	\cite{NYT2013}. 

	For the edge-attenuation parameter, $a$, in the discrete iteration  
	(\ref{eq:disc})
	there is a natural upper limit 
	given by the reciprocal of the spectral radius of each adjacency matrix $A(i)$.
	However, we note that the number of nonzeros, and hence the typical spectral radius, 
	of the adjacency matrices generally shrinks as we refine the time intervals.
	For the continuous-time ODE
	(\ref{eq:Uode})
         the principal logarithm 
   $  \log \left( I - a A(t) \right) $
     is well-defined if $1 - a \lambda_i > 0$ for all real eigenvalues $\lambda_i$ of $A(t)$,
      \cite{HigFM}.
     We note that in the case of voice calls, in the absence of tele-conferences
      $A(t)$ can always be permuted into a block diagonal structure, with nontrivial blocks 
        of the form 
 \[
  \left[
   \begin{array}{cc}
            0 & 1 \\
            1 & 0
    \end{array}
      \right],
\]
and hence this constraint on $a$ reduces to $ a < 1$.

The starting point (\ref{eq:Std}) is based on the idea of counting dynamic walks in a 
very generous sense; any traversal that uses zero, one or more edges per time point is allowed.
It is possible to take an alternative approach where other types of dynamic walk are 
counted, after downweighting. For example, we could start with the iteration 
\[
S(t+\delta t ) = \left(I+e^{-b \delta t} S(t)\right) H(A(t))^{\delta t}-I,
\]
where $H(A(t))$ is some matrix function.
Appropriate choices of $H(A(t))$ include truncated power series
of the form $I + a A(t) + a^2A(t)^2 + \cdots + a^p A(t)^p$, in which case we are
counting only those dynamic walks that use at most $p$ edges per time step in the discrete setting. 
In this generality the ODE 
(\ref{eq:Uode}) becomes
\[
 U'(t) = - b \left( U(t) - I \right) 
            + U(t)  \log \left( H( A(t)) \right). 
\]

In future work it would be natural to generalize further
to other types of dynamical system
that provide different summaries. For example,
given an evolving network $A(t) \in [0,1]^{N \times N}$ where $N$ is extremely large,
suppose
that a subset of $M \ll N $ nodes are found to be of interest.
It may then be worthwhile to study an ODE involving
$V(t)\in  \RR^{M \times M}$of the form
\[
V'(t)= P(V(t)) + Q(V(t))F(A(t)),
\]
where $P$ and $Q$ are polynomials  or other matrix valued functions, and
$F: [0,1]^{N \times N} \to  \RR^{M \times M}$ is an appropriate matrix valued
mapping. This type of system allows us to
project the interactions taking place between all $N$ individuals
onto the subset of interest,  and then calculate a measure of the resulting evolution of
behaviour at this lower dimension.

We illustrate the use of this new modelling framework on 
voice call interactions from the IEEE VAST 2008 Challenge 
\cite{DBLP:conf/ieeevast/GrinsteinPLOSW08}.
This realistic but artificial data is designed to reflect
a ficticious, controversial socio-political movement, and incorporates some  
unusual dynamic activity.
Across $400$ cell phone users over a ten day period, 
the data relevant to our experiment consists of a complete set of time
stamped calls. For each of the 9834 calls we have IDs for the send and receive nodes, a 
start time in hours/minutes
and a duration in seconds. 
We will refer to the \textbf{bandwidth} of a node as the aggregate number of seconds
for which the ID is active as a sender or receiver.
Among the extra information supplied by the 
competition designers was the strong suggestion that 
the node with ID of 200 was the ``ringleader'' in a key community.
Based on 
analyses submitted by challenge teams, we believe that 
this ringleader controls a tightly-knit subnetwork
involving nodes with IDs 1, 2, 3 and 5.
However, from day 7 onwards, these individuals appear to change their phones:
 ID 200 changes  to 300, and the others change to 306, 309, 360 and 392. 
Our aim is to show how the ODE  (\ref{eq:Uode}) is useful for dealing with this type of dynamic network.

In our test, we took $A(t)$ to be symmetric, so   
$A(t)_{ij} = A_{ji}(t) = 1$ if nodes $i$ and $j$ are conversing at time $t$, measured in seconds.
We used $b = 1/(60\times 60 \times 24) \approx 1.2 \times {10}^{-5}$, corresponding to a time downweighting of 
$e^{-1}$ per day, and a comparable value of 
$a = {10}^{-4}$ for the edge attenuation parameter.
We solved the ODE (\ref{eq:Uode}) numerically with
MATLAB's \texttt{ode23} code,
which discretizes the time interval transparently, based on its built-in 
error control algorithm.
We specified absolute and relative error tolerances of ${10}^{-4}$.
More precisely, to 
 improve efficiency, we replaced the matrix logarithm 
$\log ( I - a A(t) ) $
in 
	(\ref{eq:Uode})
 with the expansion 
$aA(t) - a^2 A(t)^2/2 +  a^3 A(t)^3/3  
 - a^4 A(t)^4/4 +  a^5 A(t)^5/5$.
Visually unchanged results were found when we increased the number of terms
in the expansion to $6$ and $7$.




Two key findings from our tests were that 
\begin{description}
   \item[(A)] without
 using any information about 
 the existence of an inner circle, 
  the dynamic broadcast/receive measures (\ref{eq:brvec})
  flag up these key nodes as being highly influential, 
   even when they are not active in terms of overall bandwidth, 
   \item[(B)] given the IDs of the inner circle, the running centrality measures
           reveal the change in the nature of the network that occurs during day 7.
\end{description}

\begin{figure}
 \scalebox{0.4}{
\includegraphics{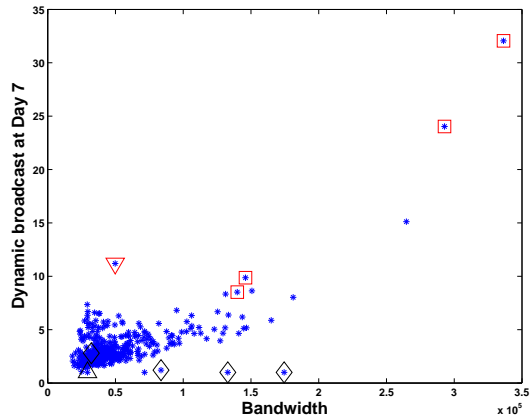}
  }
\caption{Voice call data,  end of Day 6:  for each node  we show 
dynamic broadcast measure 
on vertical axis 
 and 
bandwidth on 
horizontal axis. 
The ringleader is marked with 
 a downward pointing red triangle and the other four inner circle 
 nodes are marked with red squares.
 After day 6 these individuals are believed to change phone, the 
 ringleader's subsequent ID is 
 marked with a black upward pointing triangle, and the other inner circle IDs with black diamonds.
\label{fig.day6db}}
\end{figure}

To illustrate  point (A),
Figure~\ref{fig.day6db} takes the data up to the end of day 6, and scatter plots 
bandwidth against  dynamic broadcast, $\mb$.
(Results for dynamic receive, $\mr$, are vey similar.)
Here the ring leader, 200, is marked with a  downward pointing red triangle and the related nodes,
1, 2, 3,  and 5 are marked with red squares.
We have also marked the ringleader's follow-on ID, 300, with a black upward pointing triangle, and those for the other members, 306, 309, 360 and 392,  with black diamonds.

We see that the key nodes for day 1 to 6 are much more dominant in terms of dynamic 
broadcast than 
overall bandwidth. In particular, the ringleader node has a very modest bandwidth 
but ranks 8th out of 400 for broadcast communicability.

\begin{figure}
 \scalebox{0.4}{
\includegraphics{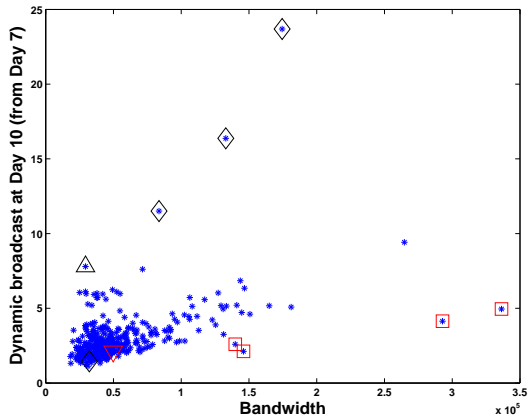}
  }
\caption{As for Figure~\ref{fig.day6db}, using data from day 7 to day 10.
\label{fig.day7db}}
\end{figure}

Figure~\ref{fig.day7db} shows the same information for the data arising 
from days 7 to 10. 
We see again that the dynamic broadcast score does a much better job of revealing the 
inner circle.
In particular, the new ID of the ringleader, marked with a black upward pointing triangle, 
has a very low overall bandwidth but ranks the 5th highest for broadcast centrality.
The `old' IDs from the inner circle, marked with red squares, continue to have 
high bandwidth, but the 
dynamic broadcast score indicates that they are no longer central players.

\begin{figure}
 \scalebox{0.4}{
\includegraphics{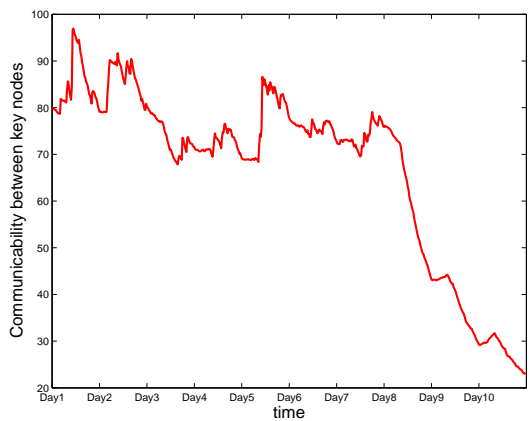}
  }
\caption{Voice call data: Dynamic communicability between the 5 key nodes (vertical axis) as 
   a function of time (horizontal axis).
\label{fig.key5}}
\end{figure}

To illustrate  point (B), Figure~\ref{fig.key5} shows the dynamic communicability between the
original IDs of the five key players, 200, 1 , 2, 3 and 5, as a function of time.
Here, at each time point,  we show the average pairwise broadcast plus receive communicability between
each pair of nodes in this group 
scaled by the average pairwise communicability between all pairs of nodes.
We see that this running measure is able to track the change in the nature of the network at 
day 7, when these individuals switch to different IDs.

In summary, this work has presented what we believe to be the first 
attempt to pass to the continuum limit 
in dynamic network centrality.
Our framework fits naturally into the context of online or digital
recording of human interactions.
The ODE setting conveniently avoids the need to discretize the network data into 
pre-defined snapshots---an approach that can introduce inaccuracies and computational inefficiencies. By defining a continuous time dynamical system, we can simulate with
off-the-shelf, state-of-the-art, adaptive numerical ODE solvers, so that time discretization   
 is performed \lq \lq under the hood\rq \rq  and in a manner that automatically handles considerations of accuracy and efficiency. In particular, in this way, we can deal 
adaptively with dramatic changes in network behaviour.
The ODE framework allows us to downweight information over time, so that a running summary 
of centrality can be updated in real time without the need to store, or take account of,  
all previous interaction history.
We focused here on the data-driven issue of monitoring
node centrality when the time-varying adjacency matrix is available. 
This has immediate applications to the issue of ranking nodes 
\cite{BGG2012},
detecting virality
\cite{IM2011}
and 
making time-sensitive strategic decisions 
\cite{BH2013}.
Further, the ODEs derived here can feed into  
the development of new network models where the evolution of centrality
is coupled to the evolution of  topology, and hence they can   
contribute to 
modelling, prediction
and hypothesis testing. 
 
PG was supported by the Research Councils UK Digital Economy Programme via 
EPSRC grant
EP/G065802/1 \emph{The Horizon Digital Economy Hub}.
DJH acknowledges support from a Royal Society Wolfson Award and a Royal Society/Leverhulme
Senior Fellowship.

\bibliography{mrefs}

\end{document}